\documentstyle[12pt]{article}
\newcommand{\eeqn}[1]{\label{#1}\end{equation}}
\newcommand{\eean}[1]{\label{#1}\end{eqnarray}}
\makeatletter
\def\citenum#1{{\def\@cite##1##2{##1}\cite{#1}}}
\def\citea#1{\@cite{#1}{}}
\makeatother

\def\npb#1#2#3{    {\it Nucl. Phys. }{\bf B #1} (19#2) #3}
\def\plb#1#2#3{    {\it Phys. Lett. }{\bf B #1} (19#2) #3}
\def\prd#1#2#3{    {\it Phys. Rev. }{\bf D #1} (19#2) #3}
\def\prep#1#2#3{   {\it Phys. Rep. }{\bf #1} (19#2) #3}
\def\prl#1#2#3{    {\it Phys. Rev. Lett. }{\bf #1} (19#2) #3}

\def\mpla#1#2#3{   {\it Mod. Phys. Lett. }{\bf A #1} (19#2) #3}

\def\ibid#1#2#3{   {\it ibid. }{\bf #1} (19#2) #3}
\def\beq{\begin{equation}}
\def\eeq{\end{equation}}
\def\bea{\begin{eqnarray}}
\def\eea{\end{eqnarray}}

\def\gtap{\ \raisebox{-.4ex}{\rlap{$\sim$}} \raisebox{.4ex}{$>$}\ }
\def\eq#1{{eq. (\ref{#1})}}
\def\eqs#1#2{{eqs. (\ref{#1}--\ref{#2})}}

\def\etal{{\it et al. }}

\begin{document}

\begin{titlepage}
\vspace*{-1.5cm}
\begin{center}

\hfill SISSA 40/94/EP
\\[1ex]  \hfill March, 1994

\vspace{4ex}
{\Large \bf Supersymmetric predictions for the inclusive  }

\vspace{-.5ex} {\Large \bf $b\to s\gamma$ decay  }

\vspace{6ex} {\bf  Stefano Bertolini } $^{b)}$
 \ \  and\ \ {\bf  Francesco Vissani } $^{a,b)}$

{\it
\vspace{1ex} a) International School for Advanced Studies, SISSA
\\[-1ex] Via Beirut 4, I-34013 Trieste, Italy
}


{\it
\vspace{1ex} b) Istituto Nazionale di Fisica Nucleare,
\\[-1ex] Sez. di Trieste, c/o SISSA,
\\[-1ex] Via Beirut 4, I-34013 Trieste, Italy
}

\vspace{6ex}
{ ABSTRACT}
\end{center}
\begin{quotation}
We study the penguin induced transition $b\to s\ \gamma$ in the
minimal N=1 supersymmetric extension of the Standard Model with
radiative breaking of the electroweak group. We include the
effects of one-loop corrections to the Higgs potential and
scalar masses. We show that the present upper and lower experimental
limits on the inclusive decay sharply constrain the parameter
space of the model in a wide range of $\tan\beta$ values. The
implications of the recently
advocated relation $|B|\ge 2$ for the bilinear SUSY soft breaking parameter
in grand unified theories are also analyzed.
\end{quotation}
\end{titlepage}
\vfill\eject

\section{Introduction}

Since the recent measurement of the first exclusive
$B\to X_s\gamma$ decay, namely \cite{CLEOex}
\beq
{\rm BR}(B\rightarrow K^*(892)\ \gamma) = (4.5\pm 1.5\pm 0.9)\times 10^{-5}
\eeqn{exclusive}
and the improved bounds on the inclusive branching ratio \cite{CLEOin}
\beq
0.8\times 10^{-4}\ <\ {\rm BR}(B\rightarrow X_s\ \gamma)\ <\ 5.4\times 10^{-4}
\eeqn{inclusive}
(at 95\% confidence level)
there has been a renewed interest in the theoretical status of
the predictions for this process in the standard model and
beyond \citea{\citenum{hewett}--\citenum{ali greub}}.

In particular it has been emphasized that the upper bound in \eq{inclusive}
can sharply constrain non supersymmetric two Higgs doublet
models \citea{\citenum{Bbook1}--\citenum{barger berger phillips}}, whereas
its impact on supersymmetric (SUSY) extensions of the model
crucially depends on the value of the ratio of the two vacuum expectation
values
$\tan\beta=v_2/v_1$ \citea{\citenum{oshimo}--\citenum{fra}}.

The measurement of
$B\rightarrow K^*(892)\ \gamma$ represents the first evidence of penguin
induced decays. The peculiarity of this loop-induced process,
$b\rightarrow s\gamma$ at the quark level,
 is that is ``dominated'' by
higher order QCD corrections, which soften
logarithmically the power-like GIM~\cite{GIM}
suppression present in the pure electroweak contribution.
This fact was first pointed out
in refs.~\cite{bbm-deshetal},
and subsequently confirmed by accurate renormalization
group (RG) analysis~\cite{QCD bsgamma}.

The logarithmic flavour-changing suppression of the QCD corrected amplitude
makes the process less sensitive to the top quark mass, but enhances
for $m_t\approx 150\ GeV$ by a
factor of 3-4 the branching ratio with respect to the purely electroweak
prediction.

The first detailed analysis of this and other $\Delta B=1$ processes
in the context  of supersymmetric extension of the SM (MSSM) has been
carried out in ref. \cite{BBMR}. In that paper the hypothesis of
an underlying grand unified theory
(GUT) was employed in order to reduce the number
of arbitrary parameters, together with the assumption of radiative breaking
of the electroweak group.
The predictions of the MSSM are there parameterized
in terms of $m_t,$
$\tan\beta$ and two independent SUSY masses.

Charged Higgs, chargino, gluino and neutralino exchange
were thoroughly studied
and the results were presented for given ranges of the four dimensional
parameter space. The analysis of
$b\rightarrow s\gamma$ and the other rare $b\rightarrow s$ transitions
considered in ref. \cite{BBMR}
showed as a general feature that, for squarks and gluinos heavier
than 100 $GeV$, gluino and neutralino induced contributions were negligible
with
respect to those induced by charged Higgs and chargino exchange.

Whereas the charged  Higgs amplitude
for $b\to s\gamma$ interferes always constructively with
the SM ($W$-induced) one, the chargino component of the amplitude
may interfere either constructively or
destructively with the previous two
depending on the region of the SUSY parameter space considered.
The region of destructive interference
in the range of $\tan\beta$ considered
in ref. \cite{BBMR} ($\tan\beta\le 8$) was observed to be confined to
a limited area of the parameter space.
It has been recently observed
\citea{\citenum{oshimo}--\citenum{garisto ng}} that for larger $\tan\beta$
the chargino contribution increases in size
allowing for branching ratios much below the SM
prediction, thus weakening substantially the potentiality of
$B\rightarrow X_s\gamma$ in constraining SUSY models.

With the present paper we want to present a detailed analysis of the
$b\rightarrow s\gamma$ transition
in the MSSM with radiative electroweak breaking,
by extending and completing
the study of ref. \cite{BBMR}.
With respect to the latter, we include here the corrections due
to the one-loop effective potential
in the determination of the minimum and shifts of the scalar masses.
In the effective potential we consider fully the contributions of
stop, sbottom and stau scalars. We have also released
the GUT scale constraint $B=A-1$ between the trilinear and bilinear
soft breaking parameters and studied the response of the model to
varying independently the two parameters.
The unification of the gauge couplings is obtained
including two-loop effects in the gauge running and
the request that bottom and tau Yukawa couplings unify
at the GUT scale (within 30\%) is imposed. For the bottom and top quarks
we include the shifts induced by QCD corrections between the pole masses and
the
$\overline{MS}$ running masses
(for a recent review see ref. \cite{arason etal}).
Considering the bottom pole mass $M_b$
between $4.6$ and $5.2$ $GeV$, we find that the requirement of tau-bottom
Yukawa unification generally pushes $M_b$ to the upper edge of the range.
We finally discuss in detail
the correlation between $\tan\beta$ and the size of the chargino amplitude
through numerical studies and analytic results.

Recent works on the topic have covered only partially the aforementioned
aspects,
either for instance
only a phenomenological approach is taken in limiting the range of the many
SUSY parameters \cite{garisto ng},
or radiative corrections to the tree-level potential
are neglected \cite{fra}.

In particular, releasing the GUT constraint $B=A-1$ turns out to have important
implications for the MSSM GUT model.
We consider interesting to examine the case $|B| \geq 2$
which is advocated
in ref. \cite{giudice roulet} as a consequence of the integration of the heavy
dynamical degrees of freedom in SUSY GUT theories.
In agreement with the ``preferred'' range
$|B| < 1$ found in ref. \cite{garisto ng},
the constraint $|B| \geq 2$ turns out
to lower the possibility of obtaining the correct vacuum and therefore reduces
substantially the accessible parameter space.

Our results are presented in figures that show the behaviour of
$b\to s\ \gamma$ amplitudes and branching ratio as a function of various
SUSY masses. We have considered the whole ranges of $\tan\beta$
allowed by the requirement of radiative breaking. The effects of the
present experimental constraints on the allowed SUSY parameter space
are shown.
We find that in the whole range of  $\tan\beta$ the GUT MSSM with
radiative breaking is already strongly constrained by the present
CLEO limits (specially in the case of $|B|\geq 2$), and that a
positive evidence for the inclusive decay might render the MSSM
quite predictive for SUSY particle searches at the
hadron colliders of the next generation.
The present analysis shows that in most of the $\tan\beta$ range
allowed, the indirect constraints on gluino, squarks and light Higgs
scalars are stronger than those obtained by direct searches at
LEP1 and Tevatron.

The paper is organized as follows.
In Sect. 2
we recall some of the basics of the minimal supersymmetric extension
of the SM, focussing on the structure of the scalar potential and of the
soft SUSY breaking sector.
In Sect. 3
we study the role of the chargino induced contribution
in the process amplitude and its characteristic dependence on $\tan\beta$.
We then analyze our numerical outcomes for the inclusive $BR(b\to s\ \gamma)$
and show their dependence on the relevant SUSY parameters.
The implications of the recent experimental observations
are finally discussed.

\newpage
\section{The MSSM}

In this section
we shall set the framework for our analysis and introduce
the necessary definitions.
After recalling the particle content and the lagrangian of the theory
we review the parameters space of the MSSM, focussing
in particular on
the structure of the model at the GUT scale, and on the soft breaking
sector.
Finally, we discuss in some
details how the renormalization group scaling of
the parameters yields us a predictive supersymmetric extension
of the Standard Model.

\subsection{Particle content and lagrangian}

In any SUSY extension of the standard
model the particle spectrum must be extended by at least
doubling the number of the Higgs fields, and introducing
a supersymmetric partner for each standard field.
One can describe conveniently the particle content of the theory
using the superfield formalism; for instance, each chiral
superfield corresponds
to a complex scalar and a
Weyl spinor, so that the
supersymmetric standard model includes
the chiral superfields $Q, U^c, D^c$,
$L, E^c,$ that extend the standard fermionic sector,
and $H_1$ and $H_2$, that extend the standard scalar sector.
These superfields transform under $SU(3)_C\times SU(2)_L \times U(1)_Y$
as follows:
$$
Q \equiv (3,2, 1/6);
\quad U^c \equiv (\bar{3},1,-2/3);
\quad D^c \equiv(\bar{3},1,1/3);
$$
$$
L \equiv (1,2,-1/2);   \quad E^c \equiv (1,1,1);
$$
\beq
H_1 \equiv (1,2,-1/2);  \quad H_2 \equiv (1,2,1/2)\ .
\label{qnumb}
\eeq

The chiral matter superfields $Q_i, U^c_i, D^c_i$ are
multiplets in
generation space; we will consider in the following the
case of three generation ($i= 1,2,3$).
Analogously each vector boson has a supersymmetric
fermionic partner.
The scalar partners of the quarks (resp. leptons) are called squarks
(resp. sleptons),
while the fermionic partners of gauge vector bosons (resp. scalars) are called
gluino, wino, zino, photino (resp. higgsino).
The spontaneous breaking of the SM gauge symmetry is obtained
by letting the scalar component of the $H_1$ and $H_2$ fields
get vacuum expectation values $v_1$ and $v_2$ respectively;
in the model
under consideration this mechanism is triggered by the running of the
relevant parameters of the Higgs potential,
as discussed in the following.
Eq. (\ref{qnumb})
shows the minimal matter content needed in any SUSY extension of
the SM, and it is the one considered in the present analysis.

We assume
that the supersymmetric standard model under consideration
is the low
energy manifestation of a Grand Unified Theory, minimally $SU(5)$.
To the degrees of freedom of eq. (\ref{qnumb})
one should add the heavy GUT superfields,
which are to be integrated away when considering the low
energy regime. Let us recall that the $SU(5)$ GUT
assumption leads to a consistent
gauge coupling constant unification in the SUSY context
at a scale $M_{GUT}\approx 3\times 10^{16}\ GeV$,
provided that the light Higgs doublet content is the minimal one (two Higgs
doublets).

The supersymmetric extension of the gauge sector
of the standard model is straightforward.
By converse the supersymmetrization of the Yukawa interactions deserves some
attention.
We build up this part of the lagrangian by constructing
the most general gauge invariant
products and sums of chiral superfields, consistent with
renormalizability (a cubic polynomial called superpotential).
This procedure leads in general to the presence of potentially dangerous
baryon (and lepton) number violating terms.
We forbid the appearance of such terms by
assuming a further symmetry in the theory, the
matter parity, or R-parity. Under this symmetry superparticle transform,
but not ordinary particles. Among else the presence of such an unbroken
symmetry implies the stability of the lightest superparticle.

When building up the superpotential one is faced with the need of
introducing two different Higgs superfields with opposite hypercharge,
to allow for mass terms for both up and down quarks and cancel
the higgsino anomaly.
A potentially dangerous $U(1)$ global symmetry in the Higgs sector
can be avoided introducing
the so called $\mu$ term in the superpotential,
which couples the two different Higgs doublet superfields.

According to the previous discussion we will henceforth refer to the following
superpotential:
\beq
W = -h_U^{ij} H_2 Q_i U^c_j + h_D^{ij} H_1 Q_i D^c_j
    + h_E^{ij} H_1 L_i E^c_j + \mu H_1 H_2
\label{supot}
\eeq
In each monomial of the superpotential
any pair of $SU(2)_L$ doublets must be contracted
with the matrix $\epsilon\equiv i \tau_2.$ The $3\times 3$ matrices
$h_U,h_D$ and $h_L$ are the complex conjugates of the usual Yukawa
matrices, namely
\beq
h_x=Y_x^*\ \ \ \ \ \ \ \ \ \ x=U,D,L
\label{conj}
\eeq
where $Y_U\ \overline{Q}_LH_2u_R$ defines the up-quark Yukawa
coupling, as it appears usually in the SM lagrangian.
The highest dimension field component of the superpotential transforms
via a total derivative under a supersymmetric transformation,
and can therefore be used to build an invariant action.

An important point on which
different kinds of  models differ is the specific structure
of the sector responsible for supersymmetry breaking.
A popular class of realistic SUSY models
is that in which the global supersymmetry
breaking is a consequence of the spontaneous breaking
of an underlying N=1 supergravity theory
(for reviews see ref. \cite{SUGRA reports}).
The locally supersymmetric lagrangian is supposed to undergo a spontaneous
breaking in the so called hidden sector,
and the effects of this breaking are communicated to the
observable sector through gravitational effects.
A renormalizable theory is obtained in
the limit in which the Planck mass goes to infinity. By doing so
we are left with a globally supersymmetric lagrangian and explicit soft
breaking terms at some GUT energy scale, which we shall call $M_X$
and for practical purposes identify tout-court with the GUT scale.
More specifically we shall consider the following gauge invariant
soft breaking Lagrangian:
\beq
{\cal L}_{soft}= -{\cal M}^2-(\hat {M}+S\ +\ h.c.)
\label{hslagr}
\eeq
where:

1) ${\cal M}^2$ is a mass term for all the scalars in the theory
\beq
{\cal M}^2 \equiv  \Sigma_i m^2_{ij} z_i^* z_j ;
\label{soft}
\eeq

2) $\hat {M}$ a mass term for the gauginos $\lambda_\alpha,$ considered
as Weyl fields
\beq
\hat{M} \equiv -\frac{M_\alpha}{2} \lambda_\alpha \lambda_\alpha;
\label{gm}
\eeq

3) $S$ is the scalar analogue of the superpotential (notice the explicit
massive parameter $m$)
\beq
S = m  \left[ -h_U^A H_2 \widetilde{Q} \widetilde{U}^c
+ h_D^A H_1 \widetilde{Q} \widetilde{D}^c
+ h_E^D H_1 \tilde{L} \widetilde{E}^c   +
B  \mu H_1 H_2 \right]
\label{trilbi}
\eeq

As we next discuss, the large number of arbitrary
parameters present in the soft breaking lagrangian of \eqs{hslagr}{trilbi}
will be drastically reduced by minimality requirements and the
GUT hypothesis.

\subsection{The low-energy minimal SUSY model}

In order to investigate the predictions of the model under consideration
for low-energy phenomenology we
must consider the renormalization group evolution of the various
parameters from the high energy scale to the scale electroweak
interactions.
This program, within the supersymmetric context,
leads to the successful prediction of $\alpha_s$ in the
framework of gauge coupling unification.
Similarly,  the tau-bottom Yukawa
unification in minimal $SU(5)$,
which depends on the yet unknown top mass and
on the ratio of the two vacuum expectation values,
can be realized in a sizable region of the parameter space
(we require unification up to 30\% correction effects due
to GUT scale thresholds and two-loop running).

In supergravity derived models the different renormalization
group evolution of the mass parameters in the scalar potential
produces the necessary conditions for the spontaneous breaking
of the electroweak symmetry. The $m_Z$ mass, seen
as a function of the various supersymmetric parameters,
allow us to reduce the size of the SUSY
parameter space (more on that in the next section).
We shall now detail our assumptions for the soft breaking
parameters.

The choice the massive parameters $M_\alpha$ and $m^2_{ij}$
is related to the supergravity model
under consideration.
For instance
either models in which
the scalar  mass terms $m^2_{ij}$ are zero at high scale (no-scale models),
or models in which the gaugino masses $M_\alpha$ are very small (models with
light gluinos),
or models in which these two parameter are related (dilaton dominated
supersymmetry breaking),
have been considered
in the literature.

Similar considerations are valid for
the choice of the
Yukawa-like parameters $h^A_{ij}$.
Our analysis will be carried out
assuming the following form of the
matrices $m^2_{ij}$ and $h^A_{ij}$:
\beq
m^2_{ij}=m^2 \delta_{ij}\ \ \ \ \ \ \ \  h^A_{ij}=A h_{ij}
\eeqn{matrix}
with $A$ a dimensionless constant.
This form
can be derived by assuming a flat K\"alher metric in the supergravity theory,
and guarantees the absence of
large flavour changing neutral currents in the scalar
sector. We will assume also that,
analogously to the gauge coupling constants,
the three gaugino masses
unify at the high energy scale as well:
\beq
M_\alpha=M\ \ \ \ \ \ \ \ \ \ \ \alpha=1,2,3
\eeqn{gaugino-init}

As a consequence, besides the yet unknown $m_t$,
we are left with five ``supersymmetric'' parameters:
\beq
A,\ B,\ M,\ m,\ \mu
\eeqn{5-param}
As a matter of fact, a relation between $A$ and $B$ holds at $M_X$
for a flat K\"ahler metric, namely $B=A-1$. This further reduces
the number of free-parameters to four.
Recently, Giudice and Roulet have made the interesting observation
that the integration of the ``heavy'' degrees of freedom of any
SUSY GUT theory with $\mu=0$ gives an effective theory
in which the original
relation $B'=A'-1$ translates to $|B|=2$ (the prime indicates the
parameters in the complete theory), and in all generality
\beq
|B|\geq 2
\eeqn{b ge2}
A calculable model dependent $\mu$ term
is then generated as a function of the original parameters.
Assuming this scenario and flat K\"ahler metric, one should then require
$|B|=2$ instead of the commonly used $B=A-1$.
Since the corresponding value of $\mu$ depends on the detailed structure of
the SUSY GUT theory, we will analyze the case $|B|=2$ leaving
$\mu$ arbitrary, and compare with the case $B=A-1$.

We conclude
this section with some considerations about
the role of the parameter $\mu.$
This parameter appears, let aside interaction
terms, in the mass matrix of the scalar particles,
of the neutralinos and of the charginos (mixed states of gauginos and
higgsinos with assigned charge).
It is important to stress that given the form of our \eq{supot}
and \eq{gm}
the chargino mass matrix reads:
\beq
- ( \widetilde{w}^- \widetilde{h}^-_1)
\left ( \begin{array}{cc}
M_2   & \sqrt{2} m_W \sin\beta \\
\sqrt{2} m_W \cos\beta & -\mu_R
\end{array} \right )
\left ( \begin{array}{c}
\widetilde{w}^+ \\
\widetilde{h}^+_2
\end{array}\right)\ \ + \ \ h.c.
\eeqn{chmatr}
where following ref. \cite{gunion haber} we define
$\widetilde w^\pm \equiv -i\lambda^\pm \equiv
-i(\lambda_1 \mp i\lambda_2)/\sqrt{2}$, and all the parameters
are taken at the weak scale ($M\to M_2$, $\mu\to \mu_R$).
It has to be stressed that abiding by the conventions of
ref. \cite{gunion haber} (which we closely follow), a minus sign
in front of the $\mu$ entries in the chargino and neutralino
mass matrices, with respect to those given in ref. \cite{gunion haber}, has
to be added to be consistent with the Feynman rules
and scalar mass matrices there given. Alternatively, one may want
to keep the plus sign in the fermion mass matrices and change the
sign of $\mu$ in the scalar mass matrices and Feynman rules.
We find more convenient to follow the first prescription.
Since the role of the chargino-squark induced amplitude is crucial
for the process we are studying it is quite important to make sure
that the relevant Feynman rules and mass matrices are derived
in a consistent way: for what matters the present analysis,
we have found a complete agreement with the
results of ref. \cite{gunion haber} up to the aforementioned $\mu$ sign
in the chargino and neutralino mass matrices.

\subsection{Radiative breaking of the electroweak symmetry}

The study of the spontaneous breaking of the electroweak group
involve a discussion of the Higgs potential and it renormalization.
Let us begin by recalling the structure of the
tree-level Higgs potential in the MSSM.
Even if we know that
the 1-loop corrections are important
(and they will be included in our analysis),
the analytic formulae that can be found at the tree-level
help in the qualitative interpretation of the numerical results.

The part of the scalar potential
involving the neutral Higgs fields
contains quadratic and quartic terms.
The massive parameters in the quadratic part are
related via the renormalization group evolution\footnote{For
a complete set of RGE's in explicit matrix form and consistent
with the present analysis see Appendix A of ref. \cite{BBMR}.
The following typos should be corrected:
in eq. (A1), $2M_1^2\to M_1^2$;
in eqs. (A6,A7),
$\widetilde{Y}_{D,U} m_Q^2 \widetilde{Y}_{D,U}^\dagger
\to \widetilde{Y}_{D,U}^\dagger m_Q^2 \widetilde{Y}_{D,U}$
and
$\widetilde{Y}_{E} m_L^2 \widetilde{Y}_{E}^\dagger
\to \widetilde{Y}_{E}^\dagger m_L^2 \widetilde{Y}_{E}$;
in eq. (A9),
$\widetilde{Y}_{x} \widetilde{Y}_{x}^{A\dagger}
\to \widetilde{Y}_{x}^\dagger \widetilde{Y}_{x}^{A}$.
}
to the massive parameters discussed in the last section,
while those of the quartic terms are directly dictated by supersymmetry:
\beq
V_0 = \mu^2_1 \vert H^0_1\vert^2 + \mu^2_2 \vert H^0_2\vert^2
- \mu^2_3 (H^0_1 H^0_2+ h.c.)
+ {1\over 8}(g^2+g^{\prime 2}) (\vert H^0_1\vert^2-\vert H^0
_2\vert^2)^2,
\eeqn{spot}
The parameters $\mu_i^2$ must be such that the scalar potential is
bounded in the direction $\vert H^0_1\vert=\vert H^0_2\vert $,
where the quartic terms gives no contribution,
and that
the configuration $<H^0_i>=0,$  corresponding to the unbroken phase,
is not a minimum of the potential.
These arguments lead to the following
relations
\beq
\mu^2_1 + \mu^2_2 >2 \vert \mu^2_3 \vert
\eeqn{stab}
and
\beq
\mu^2_1 \cdot \mu^2_2 < \mu_3^4
\eeqn{detneg}
which at the $M_X$ scale cannot be satisfied $(\mu_1^2 = \mu_2^2 = m^2 +
\mu^2)$.
However, what matters for the consistency of the model is that
\eqs{stab}{detneg} hold at $\approx m_Z$, a scale much lower than
$M_X$. In fact, due to the heaviness of the top quark
the parameters $\mu_1$ and $\mu_2$ run differently and it is
possible to satisfy both \eqs{stab}{detneg} at low energy.
The solutions depend on the chosen values for the various SUSY
parameters and in particular on $\tan\beta$: when $\tan\beta$
grows the bottom and top Yukawa coupling become more and more
similar and tend to restore the incompatibility of
\eqs{stab}{detneg} also at the weak scale.
{}From the tree-level potential analysis follows that there exist
a maximum value of $\tan\beta \approx m_t/m_b$
beyond which the electroweak breaking does not occur.
We will see that this feature is maintained even after the inclusion
of the 1-loop corrections to the potential.

It is important to recall that the renormalized parameters
are implicit functions of the Yukawa, soft breaking and $\mu$
parameters at the $M_X$ scale. We can therefore convert
relations like \eqs{stab}{detneg} into bounds on allowed
regions for the original parameters.
For instance,
the bilinear soft breaking parameter $B$ is crucial for
the determination of the electroweak vacuum.
This parameter, that enters the lagrangian through
$\mu_3^2:$
\beq
\mu_3^2|_{M_X}=-B m \mu
\eeqn{bmu}
plays a major role in drawing, through \eq{stab} and
\eq{detneg}
the regions in which the electroweak
symmetry breaking can be consistently realized.
We will see that fixing its value to $|B|\geq 2$ at $M_X$
drastically reduces the parameter space allowed by
the weak breaking.

It has been shown
that the 1-loop corrections to the Higgs potential
are important in determining the
spectrum of the physical
Higgs fields.
For instance, the MSSM tree-level prediction of a
scalar Higgs particle always lighter than the $Z$ boson is spoiled
once we take into account radiative corrections.
Furthermore, the inclusion of the 1-loop corrections
stabilize the low-energy predictions of the model
with respect to variation of the chosen renormalization scale.

The 1-loop corrected scalar potential can be written as
\beq
V_1=V_0+\Delta V
\eeqn{v1}
where $V_0$ is the tree-level potential and $\Delta V$ represents the
1-loop correction.
In the $\overline{MS}$ renormalization scheme each
particle contributes to the 1-loop potential
according to
\beq
\Delta V={1\over 64 \pi^2} {\rm Str}\left[{\cal M}^4\left( {\rm log}
{{\cal M}^2\over Q^2} -{3\over 2}\right)\right]
\eeqn{1 loop}
where ${\cal M}$ is the generalized mass matrix function of the scalar fields,
$Q$ is the renormalization scale, and Str is the supertrace,
that is the sum over the various species of particles of spin $j$ weighted
by $(-)^{2j}(2j+1)$.
Let us recall that the $Q$-dependence of the lagrangian is twofold;
the explicit $Q$-dependence in the
previous formula and the implicit $Q$-dependence
of the parameters of the lagrangian due to
the renormalization group evolution;
these two effects combine into a higher order dependence on $Q$
of the resulting effective lagrangian,
and determine the greater stability of the predictions.
We will take $Q = m_Z$,
and fully include in our numerical analysis the 1-loop
contributions of the third family of quarks, leptons,
squarks and sleptons.

An alternative way of incorporating the effects of one-loop
corrections due for instance to heavy squarks, would be
to integrate them out at some intermediate threshold ($TeV$ scale)
and consider the resulting renormalization group improved tree-level
potential \cite{olech pokorski}.
In this way the effects of large stop (sbottom)
contributions would be encoded in the different evolution of the
tree-level parameters due to presence of intermediate thresholds,
and the task
of minimization of the potential would be much simpler.
However, we choose not to follow this approach since we do not
want to assume any particle to be a priori heavier than the
present experimental bounds. On the contrary,
our aim is to test whether the constraints coming from
$b\to s\gamma$ may push some bounds further up.

\newpage
\section{Predictions for the inclusive $b\rightarrow s\gamma$ decay}

In this section we analyze the impact of the experimental information
on the inclusive $b\rightarrow s\gamma$ transition on the MSSM with
radiative breaking.

We will first devote some time to a discussion
of the individual SM and SUSY contributions
the process amplitude.
In particular, we will focus our attention on the
dependence of the chargino component on the various parameters,
and study its large destructive interference with the $W$ and $H^-$
induced amplitudes for large $\tan\beta$.

Finally, the present experimental constraints from collider and
$B$-meson physics
are used to limit the parameter space allowed for SUSY masses
below the $TeV$ scale.

\subsection{Amplitude anatomy}

For given values of $m_t$, $\tan\beta$, $B$, $M$, and $m$
($A$ and $\mu$ are then determined from the minimization of the
electroweak potential)
we numerically compute, after performing the renormalization
of the original SUSY lagrangian down to the weak scale,
the whole particle spectrum and the interaction terms
of the MSSM needed to compute the process at hand.
Then, standard QCD renormalization is used to obtain the physical
amplitude at its natural $m_b$ scale \cite{QCD bsgamma} (see also refs.
\cite{BBMR} and \cite{erice 89} for a detailed description of the
procedure used).

The $b\rightarrow s\gamma$ decay can proceed in the MSSM via five different
intermediate particles exchanges:
\begin{enumerate}
\item $W^-\ \ \ +\ \ \ $ up-quark

\item $H^-\ \ \ \ +\ \ \ $ up-quark

\item $\widetilde{\chi^-}\ \ \ \ +\ \ \ $ up-squark

\item $\widetilde{g}\ \ \ \ \ +\ \ \ $ down-squark

\item $\widetilde{\chi^0}\ \ \ \ +\ \ \ $ down-squark
\end{enumerate}

The total amplitude for the decay is the sum of all these
contributions. The complete analytic expressions for the
various amplitudes is found in ref. \cite{BBMR}.

An effective $b-s$ flavour changing transition induced by $W^-$ exchange
is the only way through which the process proceeds in
the SM. A two-Higgs doublet extension of the SM
would include the first two contributions,
while the last three are genuinely supersymmetric in nature.
Even if one might believe that the gluino exchange could be important,
due to the replacement of the weak coupling with the strong one,
it is not the case in the model under
consideration, given the present bounds on gluino and squark masses.

In Figs. 1 and 2 the relative size of the various amplitudes
to the SM one are shown. Although the figures are drawn for
given values of $m_t$ and $\tan\beta$ ($m_t=160\ GeV$
and $\tan\beta=8$) they exhibit the general features of the
various contributions for a wide range of parameters. In particular,
we observe that gluino and neutralino exchange can be neglected
in comparison with charged Higgs and chargino amplitudes
(however all the contributions are included in our final numerical results).
A thorough discussion for a qualitative understanding of the relevance of the
various component of the amplitude can be found in ref. \cite {BBMR}.

The dotted area  in the figures  corresponds to spanning the soft
breaking parameters $m$ and $M$ in the ranges $[0,800]\ GeV,$ and
$[-400,400]\ GeV$ respectively.
Both the assignments $B=A-1$ and $B=2$ are shown.
In the latter case, we show the results only for positive values of $B$
since the patterns for the corresponding negative range turn
out to be quite similar.
This can be understood by noticing that for the minimization of the tree level
potential $B\to-B$ is equivalent to $\mu\to-\mu$ (the one-loop corrections
to the potential do not sensibly spoil this feature) and that for given $B$
we always span a symmetric range in $\mu$.
We have shown
the case $B=2$ since it corresponds directly to $B=A-1$
(flat K\"ahler metric) in the original SUSY-GUT lagrangian.
In addition, it turns out to be the least
restrictive choice for the model considered in the range $|B|\geq 2$.
A discussion on the relevance of this parameter will follow.

Since gluino and neutralino contributions are numerically
irrelevant in the next figures we focus our attention
on charged Higgs and charginos by studying more in detail
the dependence of the corresponding amplitudes
on their masses and the value of $\tan\beta$.
In Figs. 3, 4 and 5 the ratios of charged Higgs and chargino
induced amplitudes with the SM contribution
are shown for
$\tan\beta=2,\ 20,\ 40,$ while the top mass is kept at $160\ GeV$.
The Higgs amplitude is shown as a function of the
charged Higgs mass, while the chargino amplitude is shown
both as a function of the lightest chargino mass and the
lightest stop mass.
No substantially different features appear by varying
$m_t$ in a few ten $GeV$ interval from our preferred value.
It should be recalled however that the maximum value
of $\tan\beta$
allowed by the radiative breaking depends on the top mass.
In the case under consideration we roughly
have $\tan\beta_{max}\approx 45$, while
for $m_t=140$ $\tan\beta_{max}\approx 40$ and
for $m_t=180$ $\tan\beta_{max}\approx 50$.
This statement is clearly dependent on the window of values of
$(m,M)$ we have considered, and on the value of $B$.
We think we have tested wide enough ranges to make the above
statements indicative
for phenomenological considerations.

Inspection of Figs. 3-5 shows that,
while the charged Higgs amplitude interferes always constructively
with the SM one (this is at the root of the sharp constraints on the
charged Higgs mass found in Higgs extensions of the SM
\citea{\citenum{Bbook1}--\citenum{barger berger phillips}}),
the chargino amplitude can give rise to substantial
destructive
interference with the SM and
$H^+$ amplitudes, becoming for large $\tan\beta$ the
dominant contribution.
This effect clearly renders the CLEO upper limit in \eq{inclusive} a
less severe constraint for the MSSM than for models in
which the negative interference is absent (see for instance
non-supersymmetric two Higgs doublet models).
In spite of that, we shall see that the full inclusion of
the present experimental limits
is already enough to exclude a large portion of the tested
parameter space.

\subsubsection{The chargino exchange contribution}

Due to the relevance of the chargino amplitude
for the present discussion, it is worth
trying to have a better understanding
of the nature of the features exhibited
by this amplitude in the previous figures.
We shall look  closely to the
formula of the chargino amplitude
for $b\to s\gamma$ derived in ref. \cite{BBMR}, to which we refer the reader
for all definitions and details:
\bea
A_{\widetilde{ \chi}^-} &=&
   - {\alpha_w \sqrt{\alpha} \over 2 \sqrt{\pi}} \
   \sum_ {j=1} ^2 \sum_ {k=1} ^6 \
   {1 \over m_{\tilde u_k}^2}                 \nonumber    \\
 & & \times   \left \{
   (G_{UL}^{jkb} - H_{UR}^{jkb})
   (G_{UL}^{*jks} - H_{UR}^{*jks})
   \left [ F_1 (x_{\tilde \chi_j^- \tilde u_k}) +
    e_U \ F_2(x_{\tilde \chi_j^- \tilde u_k}) \right ]
   \right.                                    \nonumber    \\
 & & - \left.
    H_{UL}^{jkb} (G_{UL}^{*jks} - H_{UR}^{*jks}) \
   {m_{\tilde \chi_j^-} \over m_b} \
   \left [ F_3 (x_{\tilde \chi_j^- \tilde u_k})+
    e_U \ F_4 (x_{\tilde \chi_j^- \tilde u_k}) \right ]
   \right \}
\label{phch}
\eea
where $j=1,2$ is the label of the chargino
mass eigenstates (from light to heavy) and $k=1,...,6$ is the analogous
label for the up-squarks;
the matricial couplings $G_{UL}$
arise from charged gaugino-squark-quark vertices,
whereas $H_{UL}$ and $H_{UR}$ are related to the charged
higgsino-squark-quark vertices.
These couplings contain among else
the unitary rotations $U$ and $V$
which diagonalize the chargino mass matrices.
All the quantities in \eq{phch} are defined in ref. \cite{BBMR}.

An explicit $\tan\beta$ dependence is found
in $H_{UL}$ and $H_{UR}$ where
quark Yukawa couplings are present; more precisely,
$H_{UL}$ is proportional to the down-quark Yukawa coupling,
which grows with $\tan\beta$ as $1/\cos\beta$,
whereas $H_{UR}$ contains the up-quark Yukawa coupling,
that approaches in the large $\tan\beta$ limit
a constant value ($\propto 1/\sin\beta$).
It is in fact the contribution of the third line in \eq{phch}
that determines the behaviour of the amplitude in the
large $\tan\beta$ regime (specifically the component
$H_{UL}H^*_{UR}$). Graphically it corresponds to the
Feynman diagram depicted in Fig. 6, which exhibits in terms
of squark and chargino interaction eigenstates the structure of this
component of the amplitude.
For $\tan\beta\to\infty$ the amplitude diverges, but
since we work in a perturbative scheme
we are bounded by the request of perturbativity
of the down Yukawa couplings, say $\tan\beta < 60$.

An analytic approximation of the $H_{UL}H^*_{UR}$ component
of the chargino amplitude
can be derived which shows explicitly a number
of interesting features.
Let us consider the possibility that the chargino mass matrix
in \eq{chmatr} might be approximately diagonal:
\beq
M_{\chi}\approx {\rm diag}(M_2,-\mu_R)
\eeqn{diagonal m-chi}
One can show that this approximation holds effectively when
$\ \ |M_2^2\ -\ \mu_R^2|\ \  =\ \ \ \ $
$O({\rm max}[M_2^2,\ \mu_R^2]) \gg m_W^2$ and
$M_2^2,\ \mu_R^2 \gtap m_W^2$.
It is important to notice that
these requirements, and therefore the approximation
of \eq{diagonal m-chi} are consistent with one of the eigenvalues,
say $|\mu_R|$,
being of the order of $m_W$, while the other remains much heavier.
Being the chargino mass matrix already
diagonal the approximate mass eigenvalues are simply
given by the absolute values
of the parameters $M_2$ and $\mu_R$, and the
two unitary rotations which ``diagonalize'' the
chargino mass matrix can be written as:
\beq
\begin{array}{ccl}
U &\approx & {\rm diag (sign}\{M_2\},-{\rm sign}\{\mu_R\})\ , \\
V &\approx & {\bf 1}
\end{array}
\eeqn{approx decomposition}

Using \eqs{diagonal m-chi}{approx decomposition} and the
definitions of $H_{UL,R}$ given in ref. \cite{BBMR}, we can find
a simple and instructive expression for the part of the chargino amplitude
relevant for large $\tan\beta$:
\beq
A_{\widetilde{ \chi}^-}\approx
G_F \sqrt{ \frac{\alpha}{(2\pi)^3} } K_{ts}^* K_{tb}
\left\{ \frac{1}{\sin 2\beta}
\frac{m_t}{\mu_R} \left[{\cal F}\left(\frac{m^2_{\tilde t_1}}{\mu_R^2} \right)
-{\cal F}\left(\frac{m^2_{\tilde t_2}}{\mu_R^2} \right)
\right]\right\}
\eeqn{domin ampl}
where $m^2_{\tilde t_1}$ (resp. $m^2_{\tilde t_2}$)
is the mass of the lighter (resp. heavier) stop,
$m_{\tilde\chi_1^-}=|\mu_R|$ is taken to be the lightest chargino eigenvalue
and the function
${\cal F}$ is defined to be
\bea
{\cal F}(x)
&=& \frac{1}{x}\left[F_3\left(\frac{1}{x}\right)
+e_U\ F_4\left(\frac{1}{x}\right)\right]\nonumber\\
&=& \frac{1}{6(1-x)^3}\left(5-12x+7x^2+2x(2-3x)\log x\right)
\eean{f function}

The curly bracketed term exhibit the main features
of the chargino induced amplitude for large $\tan\beta$.
Let us emphasize
three important features of this contribution.
\begin{enumerate}

\item For $\tan\beta \gg 1$ we can write $1/\sin2\beta\approx\tan\beta/2$,
showing explicitly the leading
linear behaviour of this part of the amplitude in
the parameter $\tan\beta$.

\item The value of the amplitude
is crucially dependent on the splitting of the two stop mass
eigenstates $m_{\tilde t_{1,2}},$ ${\tilde t_1}$ being the lighter stop
quark. The splitting of the two mass eigenstates
depends on the size of the L-R entry in the stop mass matrix, and
corresponds in fact
to the L-R mass insertion on the squark line in the interaction
representation of Fig. 6.
Since ${\cal F}(x)$ is a positive, monotonically
decreasing function of $x$ in the interval $x\in [0,\infty]$
(${\cal F}(0)=5/6$, $\ {\cal F}(\infty)=0$)
the term in square brackets is maximized when one of the two stop
eigenstates is light.

\item The sign of the amplitude depends directly on the sign
of the parameter $\mu_R$, that is the sign of $\mu,$ the parameter
introduced
at the GUT scale (the renormalization for $\mu$ is in fact multiplicative);
this fact means that the
region in which the chargino amplitude gives rise to a
destructive interference effect
with the other amplitudes
corresponds to the region in which $\mu$ is negative.
This behaviour was observed also in ref. \cite{lopez nanopoulos park}
but there not understood.
It is also important to notice that $\mu_R$ should be  as light as possible
($\mu_R\approx m_W$)
in order not to suppress the contribution (for large $\mu_R$ the amplitude
decreases linearly with $\mu_R$). Notice also that the amplitude
vanishes for $\mu_R\to 0$, as it should be since from the interaction state
diagram of Fig. 6 it appears that this component of the chargino amplitude
is proportional to the $\tilde h_1-\tilde h_2$
higgsino mixing, namely $\mu_R$.
\end{enumerate}

In passing it is worthwhile remarking that this component of the
chargino amplitude for the effective $b\to s \gamma$
vertex is exactly the analogue of the chargino dipole component
that in the flavour diagonal case
might dominate
the electric dipole of the elementary quarks in the MSSM,
as recently discussed in ref. \cite{bertolini vissani 1}.

\subsubsection{The role of the parameter B}

We conclude this section by remarking that if $|B| = 2$
instead of $B=A-1$ is imposed, the region of parameter
space for which
the radiative breaking of the electroweak symmetry can be realized
diminishes considerably.
We have also found that for $|B| > 2$
the allowed regions become even smaller.
As we have already remarked this observation is in agreement with the
results of ref. \cite{garisto ng}, where the authors noticed that
$B$ lying in the neighbourhood of zero represents the most favourable
case. The constraint $B=A-1$, to be imposed if one
does not require the existence of a grand unified scenario,
includes naturally this region.
A detailed and general analysis of the parameter space and
low energy particle spectrum of the model under consideration
will be matter of a forthcoming paper.

\subsection{Inclusive branching ratio: numerical results}

In Figs. 7--10
we present our numerical results for the
branching ratio of the process under consideration.
We have chosen to show them for our
preferred value of the top mass, namely
$m_t=160\ GeV$, and $\tan\beta=2,\ 8,\ 20,\ 40$ (figures
7, 8, 9, 10 respectively).
In each figure we plot the total branching ratio (which includes all SUSY
contributions) versus the three relevant SUSY masses:
that of the charged Higgs, of the lightest chargino and of the lightest
top squark (which is generally the lightest squark in the model).
The cases $B=A-1$ and $B=2$ are compared.
The horizontal solid line represents the SM result, which depends
only on the top quark mass and the value of the strong coupling
which enters through the important QCD corrections.
For the purpose of comparison with the SUSY outcomes
we show our results for a given value of $\alpha_s$, namely
$\alpha_s(m_Z)_{\overline{MS}}=0.12$. As it is widely discussed
in the literature the present experimental error on $\alpha_s$
implies an uncertainty in the predicted $BR(b\to s \gamma)$ of about
15--20\%. Even larger might be the error due to the neglect of
next-to-leading effects in the renormalization of the dipole
operator, which may be as large as 30\%
(for a recent and complete discussion
on these issues see A.J. Buras \etal in ref. \cite{QCD bsgamma}).
Here the problem is that the calculation of the
next-to-leading Hamiltonian for the $b\to s \gamma$
operator involves three-loop
mixings and their evaluation is a true formidable task.
A better understanding on the uncertainties related
to the strong renormalization of the $b\to s \gamma$ amplitude
could imply an upward or downward shift of the shaded areas
in the figures (the largest
renormalization effect is additive due to operator mixing,
and depends only on the strong coupling).
This effect becomes less relevant for large $\tan\beta$
due to the dominance of the chargino amplitude.
Keeping this in mind,
let us analyze the main features of our results.

When considering values of $\tan\beta$ of $O(1)$ (Fig. 7)
the destructive interference effect, discussed in the previous subsection,
is quite small. In this case, already the present CLEO inclusive upper bound
restricts in a sizeable way the area in parameter space allowed by
the model.
For instance, a lower bound of about $200\ GeV$ on the
charged Higgs mass is obtainable for both choices of $B$ (recall that
the corresponding LEP1 lower bound is $45\ GeV$).
However, as soon as $\tan\beta$ is of $O(10)$ and larger
the destructive interference effect
of the chargino amplitude becomes substantial and the CLEO upper
bound would not be by itself very effective in constraining the
low energy SUSY spectrum. It has to be noticed however that for
large $\tan\beta$ values and $B=2$ the allowed parameter space
for the model becomes very small and the model becomes correspondingly
quite predictive.

At any rate, a positive evidence for the inclusive decay could
be crucial in excluding large portions of the available parameter
space for the model. This can be already seen by overlapping
both the upper and lower bounds of \eq{inclusive} to Figs. 7--10.
In a more suggestive way the outcomes of imposing the full
constraint of \eq{inclusive} are shown in
Figs. 11 and 12.
In Fig. 11 the shaded areas show the implementation of the CLEO
inclusive limits
in the plane of the mass of the lightest
Higgs boson $H^0_1$ and the mass of the CP odd scalar $H^0_3$.
For a wide range of $\tan\beta$ the masses of $H^0_1$ and $H^0_3$
allowed by internal consistency of the model and the experimental
$b\to s\gamma$ bounds are higher than present LEP limits
($m_{H^0_{1,3}} \gtap 50\ GeV$,
for a recent review see for instance ref. \cite{barger phillips}).
Notice that the effect of radiative corrections on the Higgs masses
allows for the lightest Higgs boson to be heavier than the $Z$ mass.
In the present model
the allowed range for $m_{H^0_{1}}$ is bounded from above by about
$120\ GeV$, for both choices of $B$.
In Fig. 12 the corresponding regions in the gluino mass -- $\mu_R$ plane
are shown.

In conclusion, we have presented an updated study of the implications
of the recent CLEO results on $BR(b\to s\gamma)$ for the minimal
supersymmetric extension of the SM with radiative breaking
of the electroweak group. We have fully included in the analysis of the
model the leading squark (and slepton)
contributions to the one-loop effective potential
in order to make the results stable against variations of the
low-energy renormalization scale. We have imposed two loop-gauge
coupling unification and approximate tau-bottom Yukawa unification
(within 30\%). Input pole masses
($m_{\tau}$, $m_b$, $m_t$) have been related to $\overline{MS}$
running masses and approximate Yukawa matrices have been constructed at the
$m_Z$ scale using present central values for the Kobayashi-Maskawa
matrix entries \cite{PDG},
following the approach described in ref. \cite{BBMR}.

Phenomenologically interesting regions for the SUSY soft breaking
parameters have been analyzed. In particular we have discussed
the impact on the model of the constraint $|B|\ge 2$.
We have discussed in detail the
chargino contribution to the amplitude,
its specific $\tan\beta$ dependence,
which is responsible
for the large interference effects in the large $\tan\beta$ region,
and the further dependence of this interference effect
on the relevant SUSY parameters.
We have found that the combination of the upper bound on the
inclusive decay and the measurement of the lightest exclusive
channel (which leads to bounding from below the inclusive transition)
already constrains in a sizeable way the low energy
structure of the model. As a consequence, a positive experimental evidence
of the inclusive $b\to s\gamma$ decay,
hopefully together with the top discovery,
remains certainly among the most relevant (indirect) tests for
this minimal supersymmetric scenario.



\newpage
\noindent
{\large\bf Figure Captions}\\

\noindent
{\bf Figure 1.}
Ratios of SUSY induced over SM amplitudes, for $m_t=160\ GeV$,
$\tan\beta=8$ and $B=A-1$. The charged Higgs $(a)$, chargino $(b)$,
gluino $(c)$ and neutralino $(d)$ components
of the total amplitude are plotted
versus the masses of the charged Higgs, lightest chargino, gluino and
lightest neutralino respectively.

\noindent
{\bf Figure 2.}
Same as in Fig. 1 for $B=2$.

\noindent
{\bf Figure 3.}
Ratios of SUSY induced over SM amplitudes, for $m_t=160\ GeV$,
$\tan\beta=2$. Both choices $B=A-1$ $(a,b,c)$ and $B=2$ $(d,e,f)$ are shown.
The charged Higgs $(a,d)$ component
of the total amplitude is plotted
versus the mass of the charged Higgs boson which runs in the loop
together with up-type quarks,
whereas
the chargino induced amplitude is plotted both versus the mass of
the lightest chargino
$(b,e)$ and the lightest top squark $(c,f)$.

\noindent
{\bf Figure 4.}
Same as in Fig. 3 for $\tan\beta=20$.

\noindent
{\bf Figure 5.}
Same as in Fig. 3 for $\tan\beta=40$.

\noindent
{\bf Figure 6.}
The leading component of the chargino amplitude in the limit of
large $\tan\beta$ is shown in the higgsino-squark interaction basis.
The photon is attached in all possible ways. The crosses indicate
the presence of higgsino and stop mass insertions.

\noindent
{\bf Figure 7.}
The total inclusive branching ratio in the MSSM is shown for $m_t=160\ GeV$,
$\tan\beta=2$, $B=A-1$ $(a,b,c)$ and $B=2$ $(d,e,f)$, as a function of
the masses of the charged Higgs boson $(a,d)$,
the lightest chargino
$(b,e)$, and the lightest top squark $(c,f)$.
The SM prediction for $m_t=160\ GeV$ and $\alpha_s(m_Z)_{\overline{MS}}=
0.12$ is also shown (horizontal solid line) for comparison.

\noindent
{\bf Figure 8.}
Same as in Fig. 7 for $\tan\beta=8$.

\noindent
{\bf Figure 9.}
Same as in Fig. 7 for $\tan\beta=20$.

\noindent
{\bf Figure 10.}
Same as in Fig. 7 for $\tan\beta=40$.

\noindent
{\bf Figure 11.}
The scattered dot areas represent the allowed MSSM regions in the
plane of the lightest Higgs boson $H_1^0$
and the $CP$ odd scalar $H_3^0$ masses after inclusion of the
$b\to s\gamma$ bounds in \eq{inclusive}. Different values of
$\tan\beta$ are shown.

\noindent
{\bf Figure 12.}
Same as in Fig. 11, in the gluino mass~--~$\mu_R$ plane (the latter is
the $\tilde h_1- \tilde h_2$ supersymmetric mixing parameter renormalized
at the weak scale).

\end{document}